\documentclass[aps,11pt]{revtex4}
\usepackage{graphicx}

\begin{document}


\title{A Masked Photocathode in a Photoinjector}

\author{Ji Qiang}

\address{
Lawrence Berkeley National Laboratory,
          Berkeley, CA 94720, USA \\
}

\begin{abstract}
In this paper, we propose a masked photocathode inside
a photoinjector for generating high brightness electron beam.
Instead of mounting the photocathode onto an electrode, an electrode
with small hole is used as a mask to shield the photocathode from the 
accelerating vacuum chamber.
Using such a masked photocathode will make the replacement of photocathode
material easy by rotating the photocathode behind the electrode into
the hole. 
Furthermore, this helps reduce the dark current or secondary
electron emission from the photocathode material.
The masked photocathode also provides transverse cut-off to
a Gaussian laser beam that reduces electron beam emittance growth from
nonlinear space-charge effects.
\end{abstract}


\maketitle

\section{Introduction}
Photoinjector is one of the key components to provide high brightness
electron beam for next generation light sources.
The lifetime of the photoinjector depends on the lifetime of the
photocathode. The lifetime of a photocathode varies from hours to months
depending on the details of photo emissive materials and operating conditions.
For example, for the $GaAs$ photocathode used in the free electron laser
(FEL) at Jlab, the $1/e$ lifetime is about $50$ hours at an average current
of $5$ mA  
under $5\times 10^{-11}$ Torr vacuum pressure~\cite{carlos}.
For the $Cs_2Te$ photocathode at the photoinjector at the DESY
FLASH FEL facility, the lifetime is months~\cite{lederer}.
When the quantum efficiency of the photocathode becomes low, the 
photocathode needs to replaced, reactivated, or recesiated. This process
can take from hours to weeks depending on what needs to be done for
the photocathode material. For example, if only recesiation needs to be 
done for
the $GaAs$ cathode, it probably takes less than an hour by using a semi-load
lock system~\cite{carlos2}. 
On the other hand,
if reactivation has to be done, it might take much longer 
time~\cite{militsyn}.
Furthermore, the photocathode inside the vacuum chamber also 
makes significant contribution to the dark current and the secondary
electron emission inside the photoinjector due to the lower
work function of the photocathode material~\cite{schreiber,han1,han2}. 

In this paper, we propose a masked photocathode that
separates the photo emissive material from the accelerating
vacuum chamber using an electrode with a small opening hole. This
removes the dark current and the secondary electron emission from 
the photocathode material. This also protects the photocathode
from the damage of ion-back bombardment and multipacting electrons.
Furthermore, by rotating the photocathode behind the mask electrode,
a new photo emissive surface can be put into use through the opening hole.
Such a rotation can be done within minutes without taking the 
whole photocathode out. This significantly increases the usage 
lifetime of the photocathode by order of magnitude.  
Meanwhile the opening hole also provides a transverse cut-off to the 
Gaussian laser beam. This results in a more uniform transverse density 
distribution
that helps reduce the beam emittance growth from the nonlinear space-charge effects.

The organization of this paper is as follows: after the introduction, we
describe the layout of the masked photocathode in
Section 2; we study the effects of masked photocathode on
beam quality through beam dynamics simulations inside
a masked photocathode gun in Section 3;
discussions about some potential concerns are
given in Section 4.

\section{Masked Photocathode Layout}

\begin{figure}[htb]
\centering
\includegraphics*[angle=0,width=80mm]{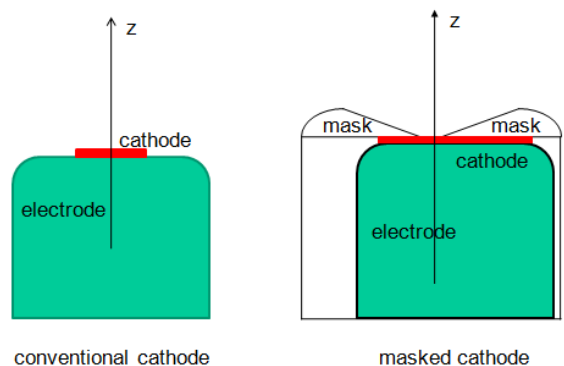}
\caption{
Side view of the conventional photocathode (left) and the masked photocathode (right).
}
\label{sideview}
\end{figure}
\begin{figure}[htb]
\centering
\includegraphics*[angle=0,width=50mm]{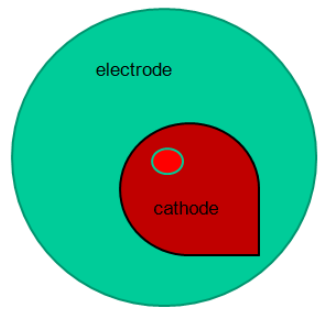}
\caption{
Top view of the masked photocathode.
}
\label{topview}
\end{figure}

Figure~\ref{sideview} shows a schematic plot of the side view 
comparison of the conventional photocathode and the masked photocathode.
In the conventional photocathode, the photocathode material is mounted 
onto an electrode made of conducting material such as $Mo$. 
The photocathode surface facing the
accelerating vacuum chamber will be exposed to the 
incident laser and also the back bombardment of ionized ions
and multipacting electrons. 
In the masked photocathode, a mask electrode with a small
hole is put in front of the photo emissive material. The size of the hole 
can be
used to control the transverse size and the uniformity of the photo electron 
beam distribution. 
The transverse
size of electron beam is normally on the order of $1$ millimeter. 
It is much smaller than the
conventional size of the photocathode 
which is on the order of a few millimeters.
This suggests that the size of the hole could be made as the same 
as the transverse electron beam size.
Such a hole helps cut off tails from a large transverse size 
Gaussian laser beam and makes the electrons out of the hole more
uniform.
The mask electrode also protects the photocathode material
from the bombardment damage of the ions and the electrons. It also
prevents the dark current and the secondary electrons being generated
from the photocathode surface. 
The size of the masked photocathode can be made relatively
large (on the order of centimeter) with an axis different from the
axis of the mask electrode.
Figure~\ref{topview} shows a top view of the masked photocathode.
By rotating and moving the axis of the photocathode behind the mask
electrode, the new photo emissive surface can be moved into the
hole for generating electrons after the depletion of the 
active photo emissive material inside the hole. Since the surface area of
the photocathode is an order of magnitude larger than the area of the hole,
this means that an order of magnitude of new photocathode
holes could be produced by simply moving the photocathode behind the mask electrode. 
This significantly shortens the time used for photocathode material 
replacement in
the conventional photocathode. To move the photocathode behind
the mask, the close contact between the photocathode and the mask electrode
is released slightly. This will avoid the damage to the photo emissive
surface due to the friction between the photocathode and the mask electrode
during the process of motion.

\section{Beam Dynamics Simulations inside a Masked Photocathode Gun}

\begin{figure}[htb]
\centering
\includegraphics*[angle=270,width=80mm]{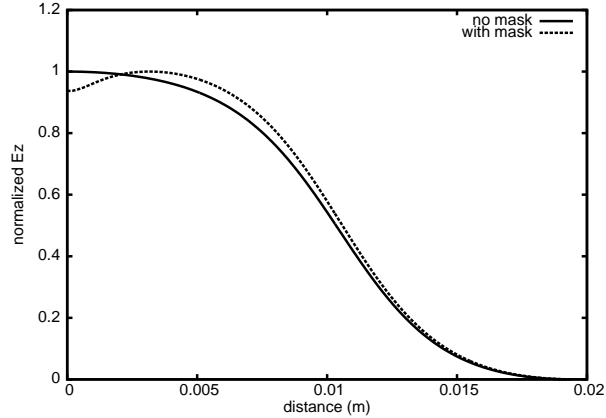}
\caption{
Normalized on-axis electric field with/without mask electrode.
}
\label{Efield}
\end{figure}
\begin{figure}[htb]
\centering
\includegraphics*[angle=270,width=80mm]{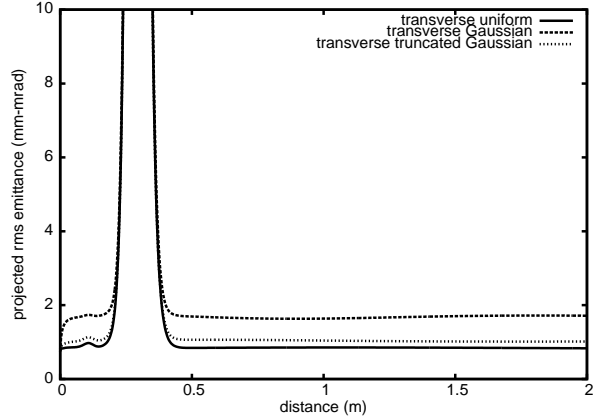}
\caption{
Transverse projected rms emittance evolution inside a photoinjector
with an initial transversely uniform distribution (solid line), 
Gaussian distribution (dashed line), and truncated
Gaussian distribution (dotted line). 
}
\label{emt1}
\end{figure}
The mask electrode provides a natural transverse cut-off
to the incident Gaussian laser beam. From the beam dynamics point of
view, this produce a more transversely
uniform electron beam that helps reduce the emittance growth of the
electron beam.
As an illustration, we use an electrostatic gap described in the
next section to study the effects of the masked cathode fields on
the electron beam dynamics through the gun.
Figure~\ref{Efield} shows the on-axis accelerating electric field
with and without the mask electrode.
The off-axis field distributions can be obtained from the derivatives
of this on-axis field following the Maxwell's equations and
the azimuthal symmetry condition. 
Without the mask electrode, the peak of the electric field is on the 
photocathode surface. With the mask, the accelerating electric field on the
photocathode inside the hole is reduced by a few percentage. 
The peak of the field is
moved downstream. This generates a transverse
field that might help focus the photo-electron beam.
\begin{figure}[htb]
\centering
\includegraphics*[angle=270,width=80mm]{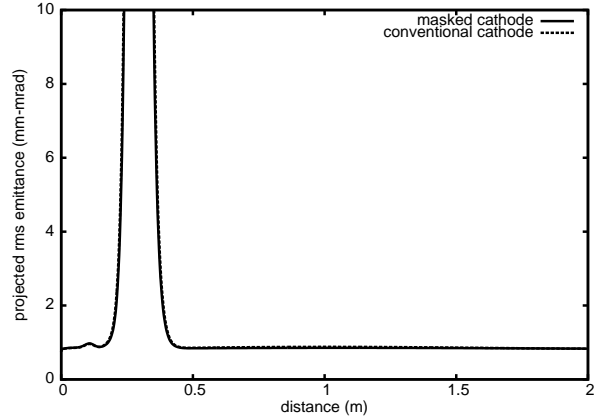}
\caption{
Transverse projected rms emittance evolution inside a photoinjector
using electric field from the masked photocathode gun and the 
conventional photocathode gun.
}
\label{emt2}
\end{figure}
Figure~\ref{emt1} shows the transverse projected emittance evolution
through the gun with an initial transversely uniform distribution,
Gaussian distribution, and truncated Gaussian distribution.
In this example, we have used the masked electric field distribution in
the Figure~\ref{Efield} with a maximum $5$ M/V field amplitude inside the gun. 
A solenoid is
placed near the exit of the gun to provide emittance compensation. 
The electron beam has $50$ pC charge with a longitudinal uniform
current distribution. The transverse rms size of the beam is $0.75$ mm. 
For the truncated Gaussian distribution, we assumed an initial beam
rms size of $1.5$ mm that is the same as the mask hole aperture radius size. 
It is seen that the final projected rms emittance with the
initial full Gaussian distribution is about a factor of two of the 
emittance with the initial uniform distribution due to the strong
nonlinear space-charge forces from the Gaussian distribution.
Using a larger laser spot size at the cathode with a mask hole 
truncates the initial Gaussian distribution. This provides a more
uniform transverse distribution for electrons out of the hole. In above
example, with $1.5$ mm rms transverse laser size, the final 
projected rms emittance of the beam is much less than the original
full Gaussian distribution and is close to the emittance
from the transversely uniform distribution. 

To check the effects of masked cathode electric field on electron
beam emittance, we also ran a comparison simulation using the masked electric
field and the conventional no-mask electric field as shown in the 
Figure~\ref{Efield} with an initial transversely uniform distribution.
The other parameters are the same as the ones used in the Figure~\ref{emt1}. 
The projected rms emittance evolution inside the gun is given in
Figure~\ref{emt2}. There is no degradation of the beam emittance from
the masked cathode field.   

\section{Discussions}

\begin{figure}[htb]
\centering
\includegraphics*[angle=0,width=80mm]{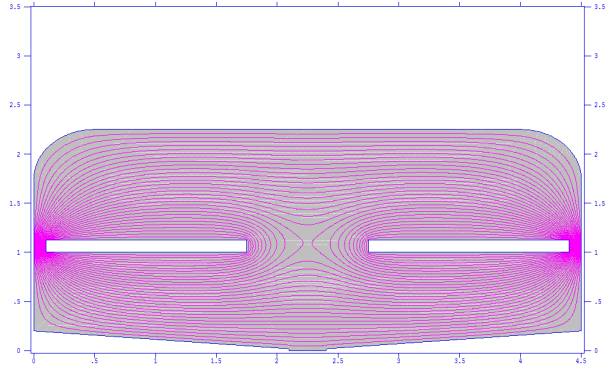}
\caption{
A schematic plot of masked photocathode together with
the anode plate.
}
\label{plate}
\end{figure}

\begin{figure}[htb]
\centering
\includegraphics*[angle=270,width=80mm]{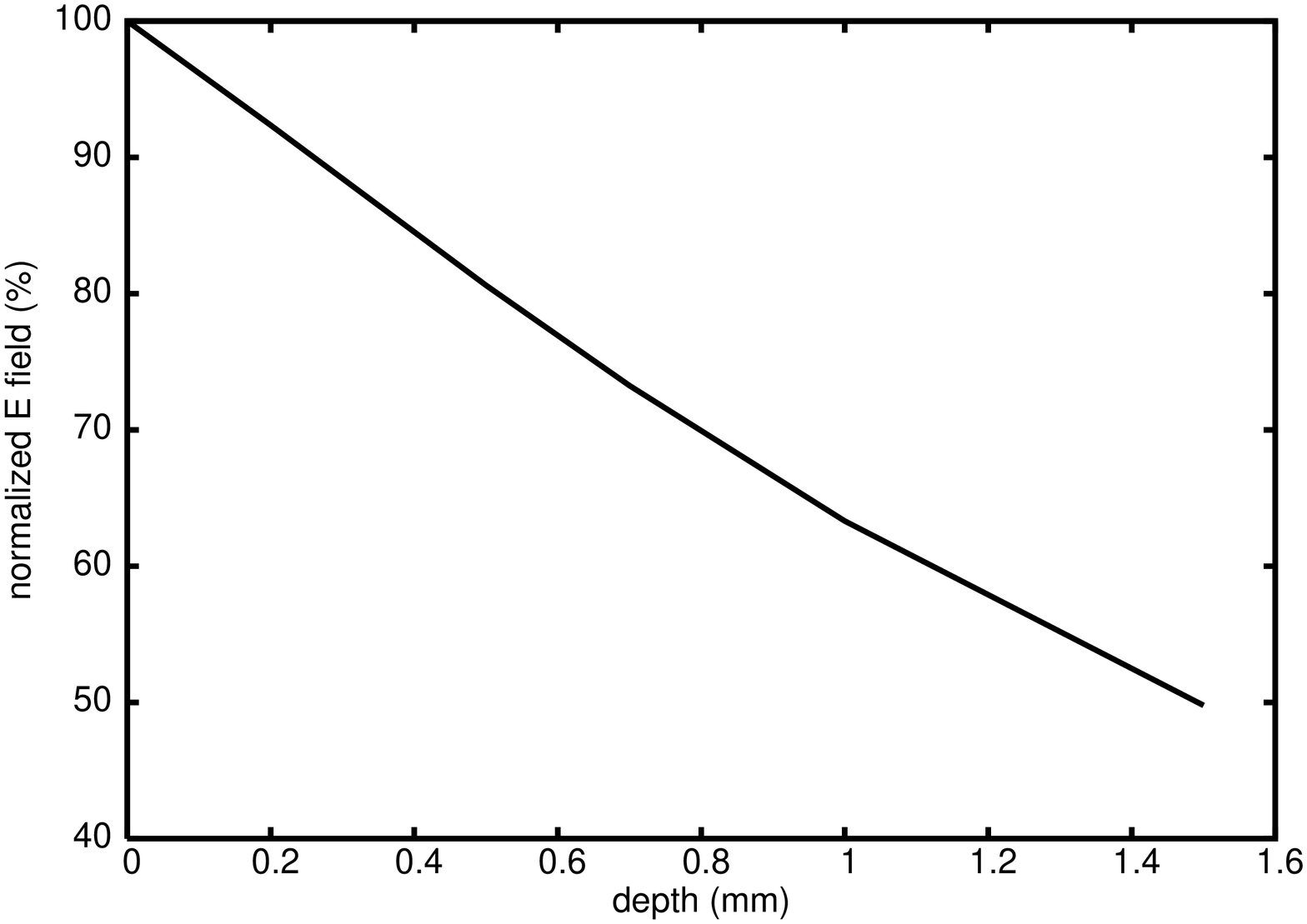}
\includegraphics*[angle=270,width=80mm]{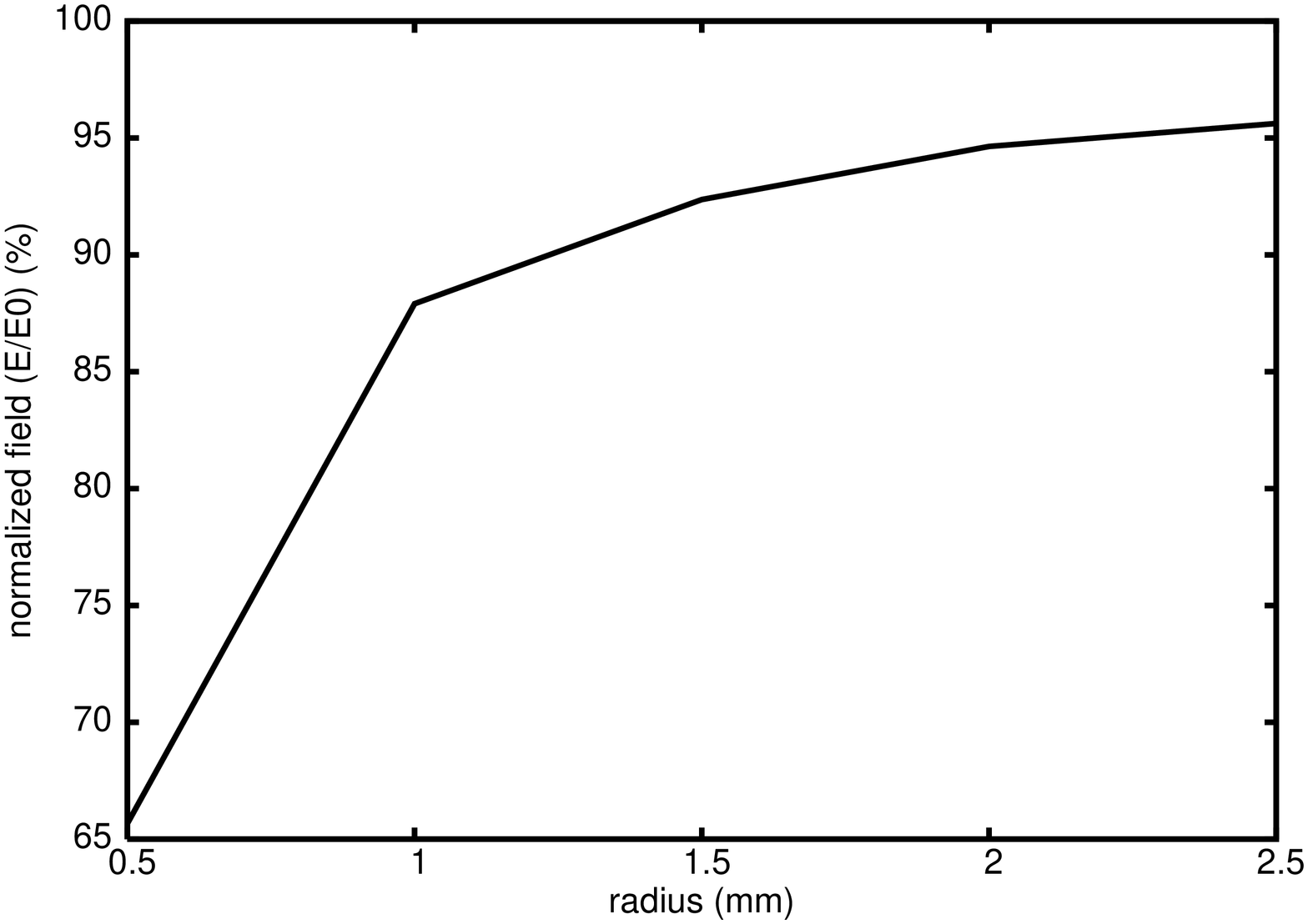}
\caption{
Electric field on the cathode as a function of the mask opening hole depth (top) 
and the mask opening hole radius (bottom).
}
\label{dep}
\end{figure}

Given the advantages of the masked photocathode described above,
there could also exist some potential challenges with this cathode.
First, the accelerating electric field on the photocathode surface
will decrease due to the shielding of the mask electrode.
To evaluate this problem, we used a simple model of two parallel plates
with a static DC voltage to study the decrease of the field on the
photocathode surface inside the hole.
Figure~\ref{plate} shows a schematic plot of computational geometry of
the two-plate structure and contours of electric potential from
the Poisson-Superfish calculation~\cite{superfish}.
Here, the mask electrode is tapered with a miminum thickness
$0.2$ mm near the hole. The radius of the hole is $1.5$ mm.
The distance between two plates is $1$ cm. 
Figure~\ref{dep} shows the photocathode surface
electric field normalized by the no-mask surface
field as a function of the mask thickness at the hole with a
fixed hole radius ($1.5$ mm) and
as a function of the hole radius with a fixed thickness ($0.2$ mm).
It appears that accelerating field at the photocathode surface goes down quickly
with the increase of the depth of the hole (i.e. the thickness of the mask).
However, as long as the thickness of the mask can be controlled
below $0.5$ mm, the accelerating field at the photocathode surface
can still reach $80\%$ of the original surface field without the mask.
Using a $0.2$ mm thick mask, the accelerating field at the
photocathode surface can attain more than $90\%$ of the original field.
With a larger hole radius as shown in the bottom plot of the figure,
the accelerating field at the photocathode surface
is larger due to the fact that
more fields will be able to penetrate into the hole.

Putting the masked electrode in front of the photocathode might result in
larger electric field on the surface of the mask electrode.
Using above numerical example, we found that the maximum electric
field on the mask surface is about $20\%$ higher
than the electric field on the photocathode surface without mask. 
Whether this increased field will cause field emission dark current 
depends on the material used for the mask electrode and the amplitude of
the electric field. Using a high work function material
helps reduce the chance of dark current from the
mask electrode surface. Coating the mask surface facing the 
accelerating vacuum chamber also helps lower the dark current.
Recent report using nitrogen-implanted silicon oxynitride film to 
coat an electrode
surface demonstrated order of magnitude 
improvement in suppressing the field emission~\cite{theodore}.

Another challenge to the masked photocathode is the
potential damage to the photocathode due to the diffusion of the
photo emissive material into the mask electrode because of the close contact
between the photocathode and the back surface of the mask electrode.
Such a problem could be overcome by using an electrode material (e.g. $Mo$)
to minimize the interaction between the photo emissive material
and the mask metal electrode. Another possible way to solve this
problem is to coat the back surface of the mask electrode with
the same photo emissive material as the photocathode.

From engineering point of view, to build the masked photocathode
might require some extra efforts and cost. This includes building
a mask electrode with very thin tip (below $0.5$ mm), and building
a photocathode supporting electrode with the capability to move
the cathode around. However, those extra efforts will be
paid back by significantly improving the lifetime of the
photocathode usage (in order of magnitude), and by improving the 
electron beam quality inside the photoinjector. Furthermore,
such a masked photocathode might also be used to generate a
flat beam for emittance exchange application by using a high aspect
ratio rectangular opening hole.

\section*{Acknowledgements}
We would like to thank Drs. D. Dowell, J. Corlett, R. Ryne, J. Staple, F. Sannibale, R. Wells,
W. Wan, M. Zolotorev for useful discussions. This research was supported by the Office of Science of the U.S. Department of Energy under Contract No. DE-AC02-05CH11231. 
This research used resources of the National Energy Research Scientific Computing Center.


\begin{thebibliography}{9}   
\bibitem{carlos}C. Hernandez-Garcia, ``Present status and future of DC
photoemission electron guns for high power, high brightness 
applications,'' in High Brightness High Power Workshop UCLA, January 14-16 2009.
\bibitem{lederer}S. Lederer et al., in Proceedings of EPAC08, Genoa, Italy,
p. 232 (2008). 
\bibitem{carlos2}C. Hernandez-Garcia et al., ``Status of Jefferson Lab FEL high
voltage photoemission guns,'' in Workshop on Sources of Polarized Electrons and High Brightness Electron Beams, Jefferson Lab, NewPort News, VA, USA, Oct. 1-3, 2008.
\bibitem{militsyn}B. L. Militsyn et al., in Proceedings of EPAC08, Genoa, Italy,
p. 235 (2008).
\bibitem{schreiber}S. Schreiber et al., in Proc. PAC 03, Chicago (USA), 2003.
\bibitem{han1}J. Han, M. Krasilnikov, K. Flottmann, Phys. Rev. ST AB, 8, 033501 (2005).
\bibitem{han2}J. Han, K. Flottmann, W. Hartung, Phys. Rev. ST AB, 11, 013501 (2008).
\bibitem{theodore}N. D. Theodore et al., IEEE Transactions on Plasma Science,
vol. 34, 1074 (2006).
\bibitem{superfish}J.H. Billen, L.M. Young, POISSON SUPERFISH, Los Alamos National Laboratory report LA-UR-96-1834 (revision January 8, 2000).

\end{thebibliography}
\end{document}